\documentstyle[aps,prl]{revtex}
\tighten
\begin{document}
\title{Mean Field Theory of
Sandpile Avalanches:\\
from the Intermittent to the Continuous Flow Regime.}
\author{V. G. Benza }
\address{ Dipartimento di Fisica, Universit\'{a} di Milano, Via Celoria 16,
Milano 20133, Italy}
\author{Franco Nori and Oscar Pla }
\address{ Department of Physics, The University of Michigan, Ann Arbor,
MI 48109-1120}
\maketitle
\begin{abstract}
We model the dynamics of avalanches in granular assemblies
in partly filled rotating cylinders using a mean-field approach.
We show that, upon varying the cylinder angular velocity $\omega$, the
system undergoes a hysteresis cycle between an intermittent and a continuous
flow regimes.  In the intermittent flow regime, and approaching the transition,
the avalanche duration exhibits critical slowing down with a temporal
power-law divergence.
Upon adding a white noise term, and close to the transition, the distribution
of avalanche durations is also a
power-law. The hysteresis, as well as the statistics of avalanche
durations, are in good qualitative agreement with recent experiments
in partly filled rotating cylinders.
\end{abstract}
\narrowtext
\twocolumn
{\it Introduction.---\/} The dynamics of granular materials is not well
understood in spite of its widespread 
scientific and technological interest\cite{Powder}.
\  Bistability, segregation, arching, hysteresis, instabilities and other
properties found in noncohesive granular assemblies make these systems
very difficult to analyze.  For over two centuries, most studies have
focused on the two extreme regimes: the fluidlike continuous flow and the
static compact state.  The intermediate quasistatic regime, with intermittent
transitions between flowing and static behavior 
has recently generated a flurry of activity including several neat
experiments\cite{Powder}-\cite{othersand} 
and novel theoretical proposals
(see, e.g., Ref. \cite{BTW87,othertheory}). 

Several experiments\cite{Chicago89}-\cite{ER}
have studied the dynamics of granular assemblies in partly filled rotating
drums.  In particular, Rajchenbach\cite{rajch90}, monitored the following
quantities:
$(i)$ the statistics of avalanches as a function
of the rotating speed of the cylinder, $\omega$, $(ii)$ the transition
from the discrete to the steady regime, including experimental evidence
of hysteresis as a function of $\omega \;$, and
$(iii)$ the dependence of the current of particles, $j$, with the angle
$\theta$ between the top layer and the horizontal.
\ It is the purpose of this work to study these issues by using a
simple mean-field model with two dynamical variables.
\ The recent studies on this subject raise many interesting questions
in granular transport, like hysteresis in $\omega$
(see, for instance, fig.~1 of ref.\cite{rajch90}),
and avalanche statistics, and this work addresses some of them.

Our mean-field approach reduces the many degrees of freedom of a granular
material in a rotating drum into two: the average velocity,
${\rm v} \ (\equiv dx/dt \equiv \dot{x})$, of
the avalanche and the average angle, $\theta$, of the granular assembly
top surface (of length $L$,  $\ 0 < x \leq L$).
\ The grains move in the ``downhill'' x--direction of
motion, which forms an angle $\theta$ with the horizontal, and the shear
between two layers with a separation $d$ is $\ \dot{\gamma}={\rm v}/d$.
\ An important feature of our approach is that the dynamics of the
model used here predicts the occurrence of avalanches as a function of
$\omega$. 
\ By slowly increasing $\omega$, we obtain the angular velocity,
$\omega_+$, at which the system undergoes a transition from the discrete
flow regime to a continuous flow (see fig.~1).  At this point, if $\omega$
is decreased, the reverse transition is observed at $\omega_-$, where
$\omega_- < \omega_+$.
\ Thus, our model exhibits hysteresis in $\omega$ during
the transition between the ``intermittent'' and ``continuous'' flow regimes.

Bagnold\cite{bagnold} extensively studied the dissipation produced by
the grain collisions in several regimes, including the quasistatic
intermittent state, finding in the latter a periodic response over time.
\ On the other hand, simple rules for the dynamics\cite{BTW87} do not
predict a periodic response over time but the absence of length and time
scales. 
\ Our model can produce distributions
in agreement with recent experiments\cite{rajch90,ER}.
\ Furthermore, when $\omega \rightarrow \omega_+$, we predict that the system
exhibits a power-law critical slowing down.

{\it Dissipation in Granular Media.---\/}
\ In granular flow, the friction force depends on ${\rm v}^2$ for large
values of the velocity. 
\ The models of Bagnold\cite{bagnold} and Jenkins and Savage\cite{savage}
correctly describe this limit.  Jaeger et al
\cite{jaeger} analyzed how the stress varies as v approaches zero,
obtaining the following form for the total
friction force:
$ 
m {\rm v}^2 / 2\lambda  + \alpha mg\cos\theta / \kappa $ where
$ \kappa = 1+ \alpha_1 {\rm v}^2 / g d \cos\theta \; ;$
which in dimensionless form 
can be written as:
$\ F=  
\beta v^2 +
\alpha \cos \theta / \kappa \; ,$
where now
$ \kappa = 1+\alpha_1 v^2 / \cos \theta  $,
$\ v = {\rm v} / \sqrt{dg} \;$,
and $\alpha$, $\alpha_1$, $\beta$, $\lambda$ are constants.
\ Here, 
the first (second) term
dominates the friction force at high (low) velocities,
and, when $v \rightarrow 0$, the second term 
represents the maximum dry friction due to rubbing forces.

{\it Dynamics.---\/}
The equation of motion
is $ \dot{v} =\sigma - F$, where $F=F(\theta,v)$ is the friction force,
and $\sigma$ is the gravitational force $(\propto \sin\theta)$.
\ If $c \equiv  \alpha \alpha_1 / \beta \; >\; 1$,
then the frictional force decreases as a function of $v$,
until a minimum is reached at $v_r$, and then increases.  We
focus on this case, $c>1$, illustrated in fig.~2(a).
\ The stationary points of $F$ determine the maximum, $\theta_m$, and
minimum (or repose) $\theta_r$, angle of stability.  Explicitly:
$\theta_m=\arctan(\alpha)$, and $\theta_r=\arctan(\alpha(2\sqrt{c}-1)/c)$.

Let us consider a stationary ($v=0$) granular assembly
with a horizontal ($\theta=0$) top surface contained in a partly filled
cylinder.  If we slowly rotate the cylinder at an angular velocity
$\dot{\theta}=\omega$, the average angle of the granular assembly surface,
$\theta$ 
grows
until it reaches the maximum angle of stability,
$\theta_m$, where the system suddenly produces
avalanches.  During this sudden transport of mass, the average velocity
of the sand grains, $v$, becomes nonzero and $\theta$
decreases until it
reaches the angle of repose, $\theta_r$,
where the system relaxes ($v$=0).  At this point the cycle repeats itself.

In order to model this dynamics, it is useful to know the time evolution
of the mean angle $\theta$.  For this purpose, let us consider the two
extreme heights (shown in fig.~2(b)) of granular vertical columns.  Since
the column heights are proportional to the mass of the columns, the rate
of mass transport is proportional to the heigth difference:
($0 < x \leq L$),
\begin{equation}
\frac{\partial h_1}{\partial t}  = - \frac{ {\rm v} }{L} (h_1 - h_2)
\hspace*{1.0in}
\frac{\partial h_2}{\partial t}  = + \frac{ {\rm v} }{L} (h_1 - h_2)  \; .
\end{equation}

The proportionality constant is the inverse of the propagation time.  The
sum of these equations give the mass conservation condition
$\ \partial_t (h_1+h_2) = 0 $, which is valid in a closed
rotating drum, while substracting them gives
\begin{equation}
\dot{\theta} \ = \ -2 \frac{ {\rm v} }{L} \tan \theta
\end{equation}
\noindent
since $h_1-h_2 = L \sin \theta $.

\ In this work, we concentrate our efforts in this $\partial_x \theta = 0$
case.  Elsewhere, we will
consider the case with local spatial variations in
$\theta(x,t)$, relevant to
describe the $S$-shaped curves seen in the continuous flow regime for
very large values of $\omega$.

We thus model the dynamics of avalanches with the following equations of
motion:
\begin{eqnarray}
\dot{v} & \ =\  & \sin\theta - \beta v^2 - \frac{ \alpha \cos\theta }
{ 1 + \alpha_1 v^2 / \cos \theta } \   =  \ \sin \theta - F(\theta, v ) \\
\dot{\theta} \ & = & \ - \rho v \tan \theta + \omega + \eta(t)
\end{eqnarray}
where $\rho = 2 \sqrt{dg} /L$, eq.(3) is Newton's equation of motion, and
the very small Gaussian noise term $\eta(t)$ simulates small fluctuations
in $\omega$ and satisfies: $<\!\eta(t)\!>=0$ and
$<\!\eta(t)\eta(t')\!>= D_0 \; \delta (t-t')\ $, with $D_0 \ll \omega$.

In the continuous flow regime, the equation for the steady state
($\dot{v}=0$, $\dot{\theta}=0$) is
\begin{equation}
\sin \theta = \beta \left( \frac{\omega}{\rho \tan \theta} \right)^2
+ \alpha \; \cos^2 \theta \left( \cos \theta + \alpha_1 \left(
\frac{\omega}{\rho \tan \theta } \right)^2 \right)^{-1} \; .
\end{equation}
This solution corresponds to any point in the branch $DC'$ in fig.~2(a)
with positive slope in $F(v)$.  Let us denote by $v_r$ ($v_m$) the
velocity corresponding to $\theta_r$ ($\theta_m$) in eq. (3)
when $\dot{v}=0$.
\ When $\omega \ll \rho v_r \tan \theta_r \ $,
the system has intermittent avalanches, and in each one of them eq.(4)
predicts that $\theta$ relaxes 
fast to $\theta_r$.
When $\omega > \rho v_m \tan \theta_m \ $,
the system is certainly in the continuous flow regime
at some point $C'$ above $C$ (see fig.2(a)).  If
$\ \rho v_r \tan \theta_r < \omega < \rho v_m \tan \theta_m $,
then the system evolves from $C$ to a point between $C$ and $D$ and
either remains there (if the stationary solution is stable at this point)
or jumps back ({\it i.e.},
$\theta \rightarrow \theta_r$ and $v\rightarrow 0$) to the $AB$ segment
and the avalanche stops.
\ If $\ \omega < \rho v_r \tan \theta_r $, then the system
certainly decays to the $AB$ segment and has intermittent avalanches
evolving through the loop $ABCD$, and so on, since $D$ is unstable.

{\it Numerical Results.---\/}
\ Figure 1 shows the time dependence of: (a) the flow velocity
$v$, (b) the angle $\theta$ of the top layer, and
(c) the angular velocity $\omega$ of the rotating drum,
obtained by numerically solving eqs.(3-4) with $D_0=0$.
\ For each avalanche, $v$ has a peak and $\theta$ has a sudden 
decrease from $\theta_m \approx 0.8$ to
$\theta_r \approx 0.65$.  The $v=0$ regions in between peaks
correspond to the loading time in between avalanches, where the angle
increases linearly in time from $\theta_r$ to $\theta_m$.
\ For clarity, only three avalanches are shown.  They are approaching the
transition point, $\omega_+$, to the continuous flow regime, and their
periods become longer close to $\omega_+$,
{\it i.e.} the system exhibits a critical slowing down.
\ The values of both $\omega_+$ and $\omega_-$ depend on $\dot{\omega}$,
and the constants $\alpha$, $\alpha_1$, $\beta$, and $D_0$ in eqs.(3-4).  The
addition of noise also reduces the hysteresis.

Notice the asymmetric form of the velocity peaks in fig.~1(a).  \
\ During a loading time, $\tau_A$, the system moves from $A$ to $B$, on
the branch $AB$ in fig.2(a).  Right afterwards, an avalanche starts and $v$
has: a sudden increase lasting a time $\tau_B$ ($B\rightarrow C$ in fig.2(a)),
a slow decrease with duration $\tau_C$ ($C\rightarrow D$), and a final
decrease lasting a time $\tau_D$ ($D \rightarrow A$).

Equations (3-4), without a random source term, and for a fixed $\omega$,
produce a periodic sequence of avalanches.  However, the real
many-body system inevitably has {\it disorder}.
\ When $D_0 \neq 0$, we observe a richer behavior (see figs. 3 and 4(b))
with broad distributions.

The statistics of the avalanche durations are shown in Fig.~3.  In order
to gain more insight into the several time scales involved in the problem,
we show the distributions, $N$, for (a) $\tau_A$, (b) $\tau_B$, (c) $\tau_C$,
and (d) $\tau_D$, $\;$ for $\omega=3 \times 10^{-4}$ and
$D_0 =10^{-8}$.

\ Figure 4(a) shows the evolution of $\tau_D$ when
$\omega \longrightarrow \omega_+$.
\ We have considered two different possibilities for the divergence:
logarithmic (left vertical axis of the inset) and power-law
(right vertical axis).
For $D_0=0$, the behavior of $\tau_D$ at the transition is found to be
\begin{equation}
 \tau_D \ \sim \ (\omega_+ - \omega)^{-p}
 \hspace*{1.0in}
{\rm for \ \ \/} \omega < \omega_+ \; .
\end{equation}
\ Temporal critical slowing down behavior has also been found in bistable
optical systems\cite{benza}.
\ From eq.(6) and recalling that $\omega + \eta$ is a Gaussian random
variable, one obtains that $N(\tau_D) \sim \tau_D^{-p} $.  This
is checked numerically in fig.~4(b) where we have obtained the distribution
of $\tau_D$ for fixed $\omega = 4.6 \times 10^{-3}$,
which is slightly below $\omega_+(D_0=0)$, 
with $D_0=10^{-8}$.

In the intermittent flow regime, our dynamical model exhibits an attractive
limit cycle around an unstable fixed point.  For $\theta < \theta_m \;$,
$v=0$ is a stable attractor manifold.  In the continuum flow regime, the
dynamical system is in an attractive fixed point with nonzero velocity.
\ Here, a harmonic approximation to Newton's
equation (eq.(3)) with $\dot{v}=0$ gives
\begin{equation}
v-v_r \simeq (\theta-\theta_r)^{1/2}
\end{equation}
for the relationship between the mass current flow and the angle, for
$\omega \geq \omega_- $.  This result
is consistent with experimental results by Rajchenbach\cite{rajch90}.

{\it Limitations.---\/}
Our mean-field equations have only two degrees of freedom for modeling
an extremely complex system.
Thus, it might need to be extended in order
to provide a more complete description of the system.
\ The beauty of the model lies precisely in its simplicity and in the fact
that it can naturally describe several dynamical features observed in
experiments in granular media.  

{\it Conclusions.---\/}
We have studied a mean-field model of granular assemblies driven to the
threshold of instability, where they produce avalanches.
\ From this simple model, we obtained the distributions of durations of
avalanches, the transition between the two flowing regimes, and the
hysteresis between them in a natural manner, and with a good qualitative
agreement with results recently obtained using experiments in rotating
cylinders.  Furthermore, when $\omega \rightarrow \omega_+$, the avalanche
durations exhibit a critical slowing down with a temporal power-law
divergence.

We thank the assistance of E. Todd.  This work was supported in part
by the INFN I.S. M19 (VB), NSF grant DMR-90-01502 (FN) and the
Spanish Ministry of Education (OP).

\begin{figure}
\caption{Shows the time dependence of: (a) the flow velocity $v$,
(b) the angle $\theta$ of the top layer, and
(c) the angular velocity $\omega$ of the rotating drum.
\ Throughout this work, we use the parameters values:
$\alpha=1$, $\alpha_1=2$, $\beta=1/2$, and $\rho=10^{-3}$.
\ For these ramping rates, we obtain
$\omega_+=4.76 \times 10^{-4}$ and
$\omega_-=4.66 \times 10^{-4}$,
for the $D_0=0$ deterministic case.}
\end{figure}

\begin{figure}
\caption{(a) Schematic plot of the friction force versus velocity
$v$ for the case $c>1$.
\ An initially flat ($\theta=0$) and stationary ($v=0$) granular surface
is tilted with an angular velocity $\omega$, thus having the following
sequence of values for the friction force:
$O \rightarrow A \rightarrow B$, from the $F=v=0$ origin, $O$.  The velocity,
$v$, is zero in the stick-slip
case with infinite slope for the $OAB$ segment.
\ In the $AB$ branch, the loading time, $\tau_A$, satisfies
$<\!\tau_A\!> = (\theta_m-\theta_r)/\omega$.
\ When the maximum angle of stability is reached in $B$,
an avalanche ($v \neq 0$) starts and the system jumps to $C$.
\ Different situations, described in the text, can now develop according to
the value of $\omega$.  (b) Schematic diagram of a rotating drum indicating
the extreme heigths, the mean downhill velocity $v$, and the mean
angle $\theta$.}
\end{figure}

\begin{figure}
\caption{Statistics of the avalanche durations.  Distributions, $N$, for
(a) $\tau_A$, (b) $\tau_B$, (c) $\tau_C$, and
(d) $\tau_D$, for $\omega=3 \times 10^{-4}$ and
$D_0=10^{-8}$.}
\end{figure}

\begin{figure}
\caption{(a) Time evolution of $\tau_D(\omega)$ when
$\omega \longrightarrow \omega_+(D_0=0)$.  This critical slowing down with
a power law divergence is shown in the
inset by the dashed line corresponding to the right vertical logarithmic
axis.  For comparison purposes, the solid line in the inset, corresponding
to the left vertical linear axis, has a small curvature
because of the small value of the power ($p \simeq 0.34$).
\ (b) Distribution of $\tau_D$ for
$\omega = 4.6 \times 10^{-4}$,
which is slightly below $\omega_+(D_0=0)$.  
\ The inset shows the same data in semi-log and log-log scales.  }
\end{figure}
\end{document}